\pdfoutput=1

\documentclass[twocolumn]{article}

\usepackage{graphicx}
\usepackage{amssymb,amsfonts,amsmath}

\newcommand{\comment}[1]{}

\def\VDS{$V_{DS}$}
\def\VQPCa{$V_{QPC1}$}
\def\VQPCb{$V_{QPC2}$}

\def\VG{$V_g$}
\def\VTG{$V_{TG}$}

\def\fedge{$f_{edge}$}
\def\fR{$f_R$}
\def\fT{$f_T$}

\begin{document}

\title{The Role of Interactions in an Electronic Fabry-Perot Interferometer Operating in the Quantum Hall Effect Regime}

\author{Nissim Ofek, Aveek Bid, Moty Heiblum, Ady Stern, Vladimir Umansky \& Diana Mahalu \\
 \\
Weizmann Institute of Science, Rehovot 76100, Israel}

\maketitle

\begin{abstract}
Interference of edge channels is expected to be a prominent tool for studying statistics of charged quasiparticles in the quantum Hall effect (QHE) \cite{STE2008, CHA1997}.  We present here a detailed study of an electronic Fabry-Perot interferometer (FPI) operating in the QHE regime \cite{CHA1997}, with the phase of the interfering quasiparticles controlled by the Aharonov-Bohm (AB) effect.  Our main finding is that Coulomb interactions among the electrons dominate the interference, even in a relatively large area FPI, leading to a strong dependence of the area enclosed by the interference loop on the magnetic field.  In particular, for a composite edge structure, with a few independent edge channels propagating along the edge, interference of the outmost edge channel (belonging to the lowest Landau level) was \emph{insensitive} to magnetic field; suggesting a constant enclosed flux.  However, when any of the inner edge channels interfered, the enclosed flux \emph{decreased} when the magnetic field \emph{increased}.  By intentionally varying the enclosed area with a biased metallic gate and observing the periodicity of the interference pattern, charges e (for integer filling factors) and e/3 (for a fractional filling factor) were found to be expelled from the FPI.  Moreover, these observations provided also a novel way of detecting the charge of the interfering quasiparticles.
\end{abstract}

A considerable amount of work has been focused in recent years on interference of quantum particles,
aiming at understanding `single-particle' as well as `correlated-particles' physics.  Experiments were designed to measure the quantum coherence time \cite{NAY2008}; to determine phase of the scattering amplitudes \cite{SCH1997}; to test entanglement between a pair of quantum particles \cite{NED2007}; and, more recently, to probe their charge and coherence properties integer and fractional QHE regimes (IQHE \& FQHE, respectively) \cite{WIL2009, CAM2007, ROS2007}.  However, even though the statistics of identical quantum particles upon exchange is at the core of many physical phenomena, experiments to directly observe this intrinsic property are scarce \cite{WIL2009,CAM2007}.  Recent interest in topological quantum computation \cite{KIT2003,NAY2007,DAS2005} led to a series of theoretical proposals aimed at demonstrating the statistics of quasiparticles in the FQHE \cite{BON2006}.  The ideas are primarily based on performing interference in an electronic version of a Fabry-Perot interferometer (FPI) \cite{STE2008,CHA1997} or in a similar Mach-Zehnder interferometer (MZI) \cite{JI2003,LAW2006}; with distinct `fingerprints' of the interfering quasiparticles.  The MZI, while being more difficult to fabricate, is a `true two-path' interferometer.  On the other hand, in the FPI, which is a simpler device; many trajectories contribute to the interference.  Understanding the physics governing these interferometers is crucial if they are to be used to probe the statistics of quasiparticles.

We performed a detailed study of an electronic FPI operating in the IQHE and the FQHE regimes.  Different from previous works, our devices allowed the study of interference of any of the propagating edge channels; at integer and fractional filling factors; for a few areal carrier densities; the dependence on magnetic field; as function of the FPI area; and at different coupling strengths to the reservoir leads.  Previous works \cite{CAM2007, WEE1989,ZHA2009} reported on more restricted phenomena (mostly on interference of the innermost channel), hence, missing some of the more striking effects we report here.  Moreover, our measurements in the FQHE regime are strikingly different from those of Goldman's group \cite{CAM2007}, which have all along been difficult to interpret \cite{CHE2009}.  Our data aims at attaining a rather complete picture of the behavior the FPI in the QHE regime, with a particular focus on the role of electron-electron interaction that seem to dominate the interference.

\section{Experimental setup}
The electronic FPI, discussed in some detail first by Chamon et al. \cite{CHA1997}, was realized in a form of a Hall bar perturbed along the current flow by two constrictions; each formed by a biased split-gate quantum point contact (QPC), with a plunger gate that controlled the area of the FPI (shown in Fig. 1). Current flows chirally in the form of distinct edge channels, with the lower right-moving channel biased (at voltage \VDS) relative to the upper left-moving one (kept at
\begin{figure}[h!]
\label{fig:setup}
\centerline{\includegraphics*[viewport=1.5in 2.0in 6.8in 9.5in,width=82mm]{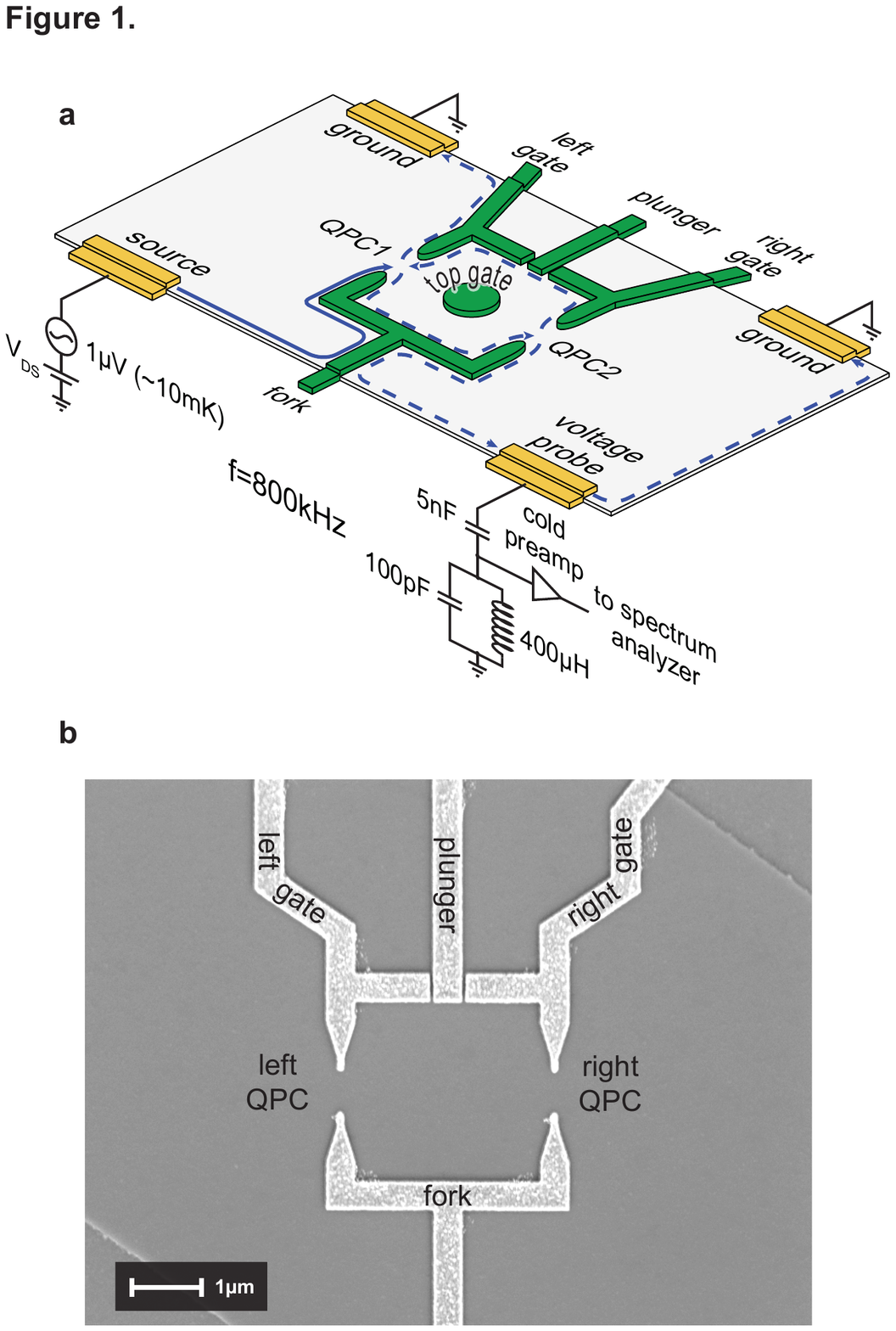}}
\caption{Measurement circuit and the Fabry-Perot interferometer.  (a) A cell of 2DEG confined by two quantum point contacts (QPCs) constrictions, with transmissions $t_i$ (reflection $r_i$=1-$t_i$) determining the confinement of the electrons.  Measurements are carried out with the bulk in a quantum Hall plateau, with a constant voltage source $\sim$1$\mu$Vp-p at 800kHz.  The voltage measured at the voltage probe is proportional to the transmission probability through the interferometer.  An optional `top gate' covers a portion of the cell (see Fig. 4 \& 8) (b) SEM image of the FPI where the bright regions are the metallic gates.  The cell area, after taking into account a depletion width of 250nm, is about 4.4$\mu$m$^2$.}
\end{figure}
ground potential).  When the bulk is in a QHE state, the two-terminal conductance, in the absence of the constrictions, is quantized at $\nu\cdot e^2/h$, with $\nu={n\phi_0}/{B}$ being the filling factor, where $n$ is the areal electron density, $\phi_0=h/e$ the flux quantum, $e$ the electron charge, $h$ the Planck constant, and $B$ the magnetic field.  Each constriction induces backscattering from a forward propagating edge channel to a backward moving one.  Circulating edge currents in the FPI acquire a phase determined by the AB effect (proportional to the enclosed magnetic flux) and the number of enclosed quasiparticles.  The AB phase can be controlled by either changing the magnetic field or the area of the interferometer (via biasing the plunger gate).

Our interferometers were fabricated on heterostructures embedding 2DEG with a few different areal electron densities,
$n=$5.6$\times$10$^{10}$cm$^{-2}$, 8.6$\times$10$^{10}$cm$^{-2}$ and 2.0$\times$10$^{11}$cm$^{-2}$, all having low temperature mobility in excess of 10$\times$10$^6$cm$^2/$V-s. Measurements were done on two different sizes FPIs, one with an area $\sim$4.4$\mu$m$^2$ and one with an area $\sim$17.6$\mu $m$^2$ , with plunger gate width $0.4\mu m$.  Here we present data mainly taken with the smaller FPI (unless otherwise specified), fabricated on the wafer with electron density 8.6$\times$10$^{10}$cm$^{-2}$ (see Fig. 1b). Samples were cooled in a dilution refrigerator to an electron temperature $\sim$10mK (verified by temperature sensitive shot noise measurements) with conductance measurements taken at a typical excitation voltage of about 1$\mu$Vp-p at 800kHz.  The parameters of the system are as follows: the area of the interferometer, $A$; the magnetic field, $B$; the QPCs split-gate voltages, \VQPCa~and \VQPCb; the plunger gate voltage, \VG; the bias on the source, \VDS; the electron temperature, $T$; and in some devices the voltage applied to a top-gate, \VTG.

The cell of the FPI, is defined by depleting the electrons underneath the metallic gates that surrounds it while assuring that the two constrictions stay open.  Then, the magnetic field is tuned to the approximate center of a conductance plateau; with the filling factor in the FPI cell similar to that in the bulk (verified by multi-terminal Hall measurements across the interferometer).  The constrictions are then being carefully pinched (via the split-gate voltages, \VQPCa~\& \VQPCb), resulting in transmissions of the interfering edge channel $t_1$ and $t_2$, respectively.  We denote the number of fully reflected edge channels, belonging to the highest Landau levels, being also the number of circulating edge channels within the FPI cell, by \fR, and the number of fully transmitted edge channels, belonging to the lowest Landau levels, by \fT.  Together with the interfering channel the total number of channels is \fedge=\fT+1+\fR; being determined by the bulk filling factor $\nu$ (for IQHE, \fedge=$\nu$).  We notate each configuration as ($\nu$, \fT)$\equiv$(bulk filling factor, number of fully transmitted edge channels).  The two terminal differential conductance ${\partial I}\left/{\partial V_{DS}}\right.$ was measured as function of various parameters, ${\partial I}\left/{\partial V_{DS}}\right.\left(B, V_g, f_T, t_1, t_2\right)$ , from which the periodicities of the AB oscillations were extracted, $\Delta B\left(B, V_g, f_T, t_1, t_2\right)$ ($\Delta B\ll B$) and $\Delta V_g\left(B, V_g, f_T, t_1, t_2\right)$.  We selected to present here only a subset of our results, which, however, portray a rather complete picture of the behavior of the FPI.

In the IQHE regime, interference was measured for all possible configurations ($\nu$, \fT), with \fT=$1..\nu-1$ and $\nu=1$ through 6 and at different settings of transmissions $t_1$ and $t_2$.  As expected, for $t_1$, $t_2\ll1$ (weak coupling to the leads) multiple reflected trajectories within the FPI interfered, and the two terminal conductance peaked sharply as function of flux; while at high transmissions $t_1$ and $t_2$, the conductance was almost sinusoidal as function of flux.  However, and this is important to note, the periodicities of the interference were independent of the transmissions.

\section{Experimental results}
We start with $(1, 0)$, namely, a single propagating edge channel (the spin resolved first Landau level), with an
oscillating conductance as function of \VG, with periodicity $\Delta V_g=$2.35mV (inset in Fig. 2).

\begin{figure}[h!]
\label{fig:b_indep}
\centerline{\includegraphics*[viewport=1.2in 3.5in 7.0in 8.1in,width=82mm]{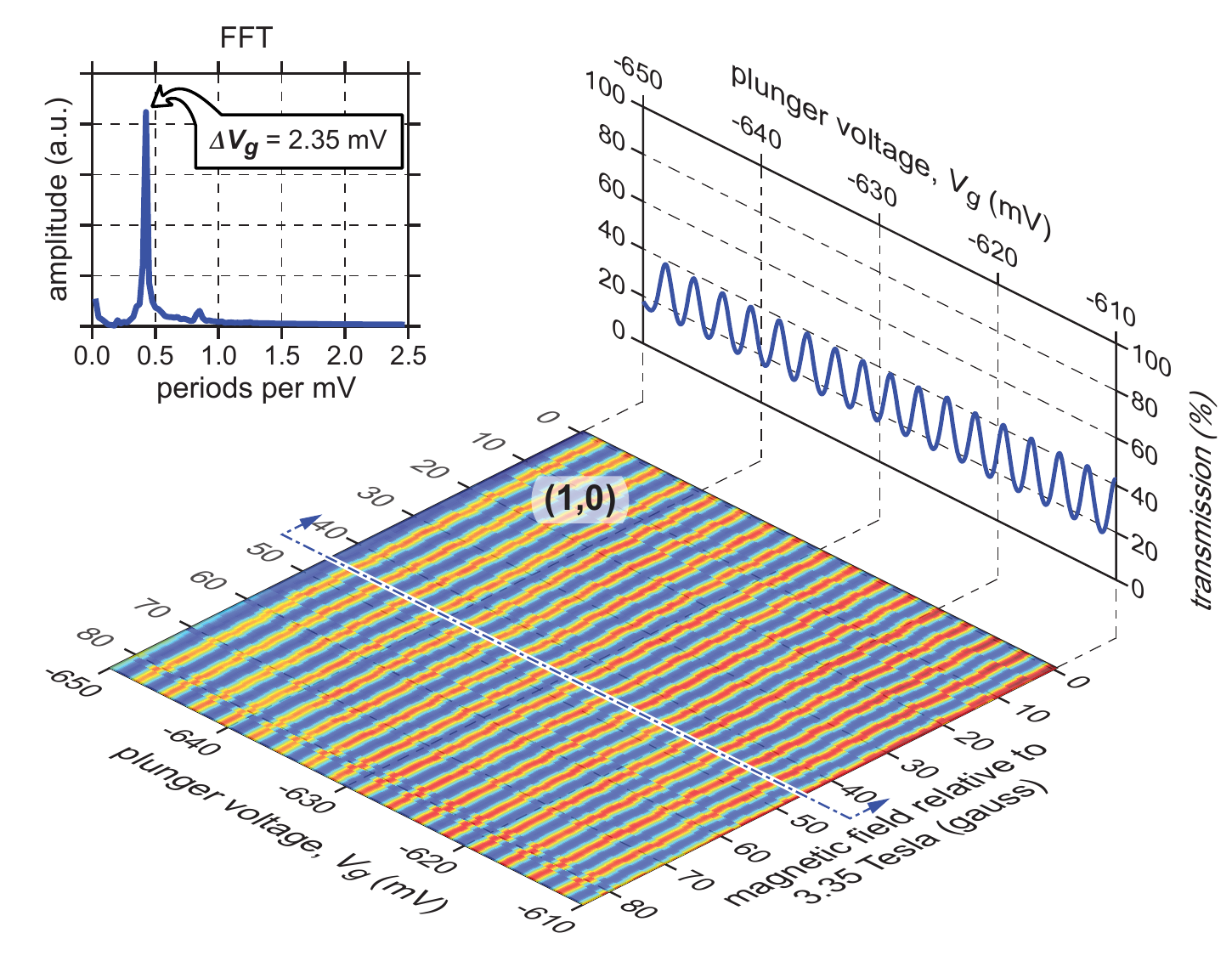}}
\caption{Interference of the outer edge channel at $\nu$=1 and \fT=0.  While independent of magnetic field, the transmission oscillates as function of plunger voltage, \VG.  The upper right panel shows a cut in the 2D $B$-\VG~plot.  The upper left panel shows FFT of this cut with a periodicity of 2.35mV, corresponding to the expulsion of one electron per period.  Note that a weak second harmonic is also observable - corresponding to two fully rotating trajectories around the interferometer.}
\end{figure}

Remarkably, the conductance in this case (and universally for all cases of \fT=0) was found to be independent of magnetic field for many added flux quanta (the magnetic field corresponding to one flux quantum threading 4.4$\mu$m$^2$ is $\sim$9.8gauss).  So here we had an extraordinary situation where AB oscillations were responsive to a change in area but not to a change in the threading magnetic field.  In contrast, profound oscillations as function
of both $B$ and \VG~were always observed for \fT$>$0, with a typical behavior for case $(3, 2)$ shown in Fig. 3.  The periodicity with \VG~was found to depend only on \fT~(in both, the IQHE and FQHE regimes), increasing when \fT~increased.  Unexpectedly, the slope of the constant phase lines in the $B$-\VG~plane is counter intuitive: \emph{an increase} in $B$ \emph{must be accompanied by an increase} in \VG~in order to keep the phase constant (as was also observed in Ref. \cite{ZHA2009}).  This suggests that an increase in $B$ leads to a decrease of the enclosed flux - an opposite behavior to that observed before in the electronic MZI \cite{JI2003}.  An exception to this behavior, observed in our large FPI, will be discussed separately.  In the example in Fig. 3a [case $(3, 2)$], the dependence of the transmission on $B$ and on \VG~was sinusoidal for large $t_1$ and $t_2$; becoming an array of periodic sharp peaks when the constrictions were pinched (Fig. 3b), however, the periodicities did not change.

\begin{figure}[h!]
\label{fig:t_indep}
\centerline{\includegraphics*[viewport=0.9in 2.5in 7.2in 8.9in,width=82mm]{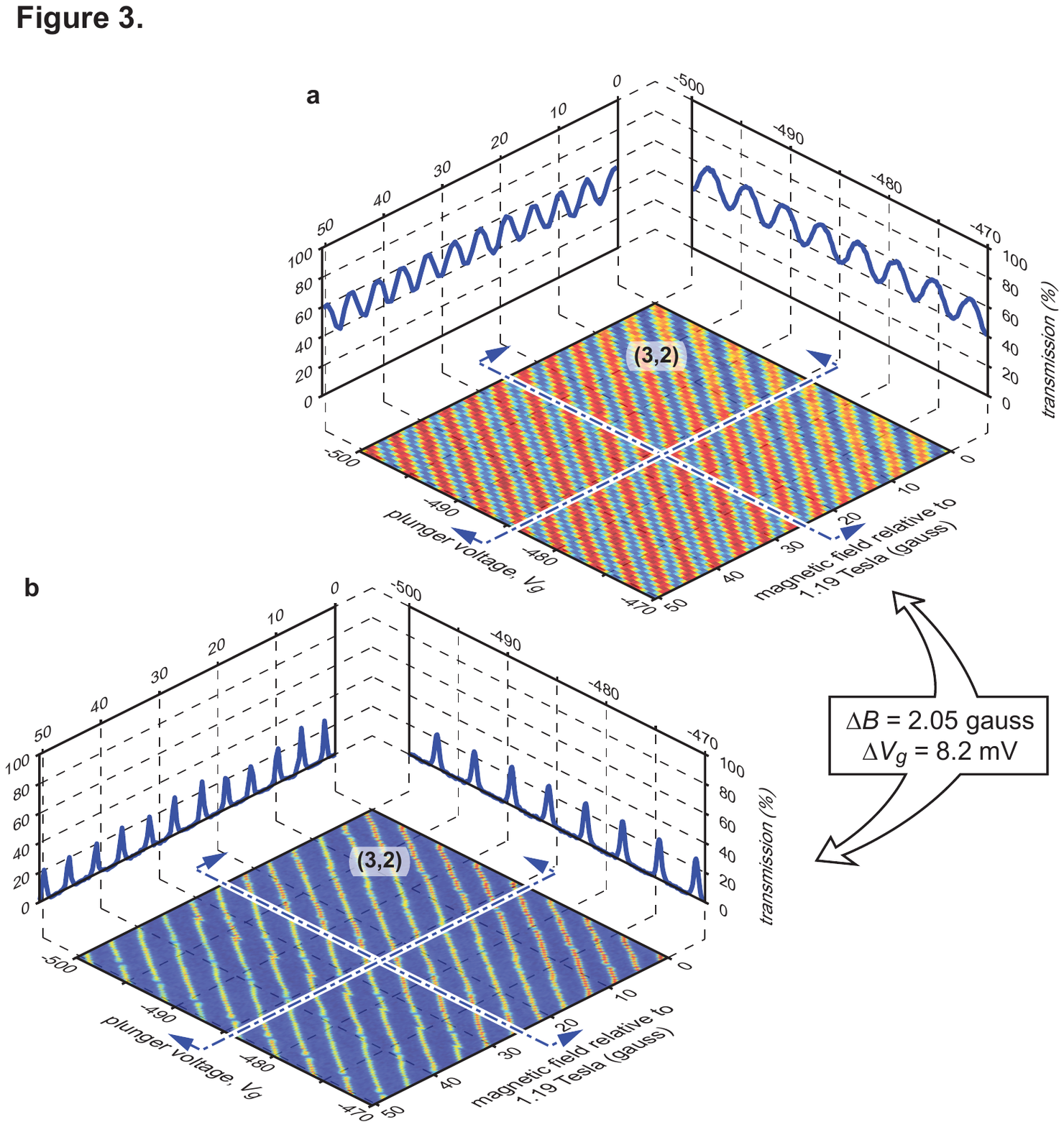}}
\caption{Interference of an inner most channel at $\nu$=3 and \fT=2. (a) Color scale plot of interference in $B$-\VG~plane with relatively open constrictions ($t\approx0.7$), leading to a sine-like dependence of the transmission on $B$ and \VG.  The slope of the constant phase lines is opposite to the expected one if interactions between the electrons are negligible. (b) Color scale plot of interference in $B$-\VG~plane with relatively pinched constrictions ($t\approx0.3$), leading to sharp periodic peaks of the transmission.  Note that the periodicities in both cases are identical.}
\end{figure}

Similar conductance oscillations were observed as function of a voltage \VTG~applied to an added small top-gate ($TG$) in the center of the FPI cell, with area $\sim$0.75$\mu$m$^2$ and circumference $\sim$3$\mu$m (see Fig. 4a).  Again, the periodicities in \VTG~for $(1, 0)$ and $(1/3, 0)$, are identical - as summarized in Fig. 4b.  For \VTG$>$-135mV, the area under the top-gate is being only partly depleted, and the periodicity observed is $\Delta$\VTG$\sim$0.23mV.  A trivial two-plate capacitor model, with $C_{TG}\sim$0.66fF calculated for our specific device, suggests an expulsion of one electron in a single period.  For \VTG$<$-135mV, the 2DEG under the top-gate is fully depleted and the depletion region extends away from the top-gate.  The capacitance of the top-gate to the 2DEG slowly increased with gate voltage, starting at a periodicity of 0.23mV and increasing to $\sim$0.6mV. The periodicity measured with a plunger gate for the \fT=0 cases if about 2.3mV. Dividing it by the ratio between the top-gate length and its own ($\sim$0.4$\mu$m$^2$) gives about 0.30mV. This is consistent with $\Delta$\VTG~measured at beyond depletion voltages. A rather interesting behavior is observed in the \VG-\VTG~plane when looking at the slope of the constant phase lines in Fig. 4c; both gates, plunger and top, lowered the area when biased negatively.  We address this issue later.

\begin{figure}[h!]
\label{fig:antidot}
\centerline{\includegraphics*[viewport=0.9in 3.5in 7.5in 8.1in,width=82mm]{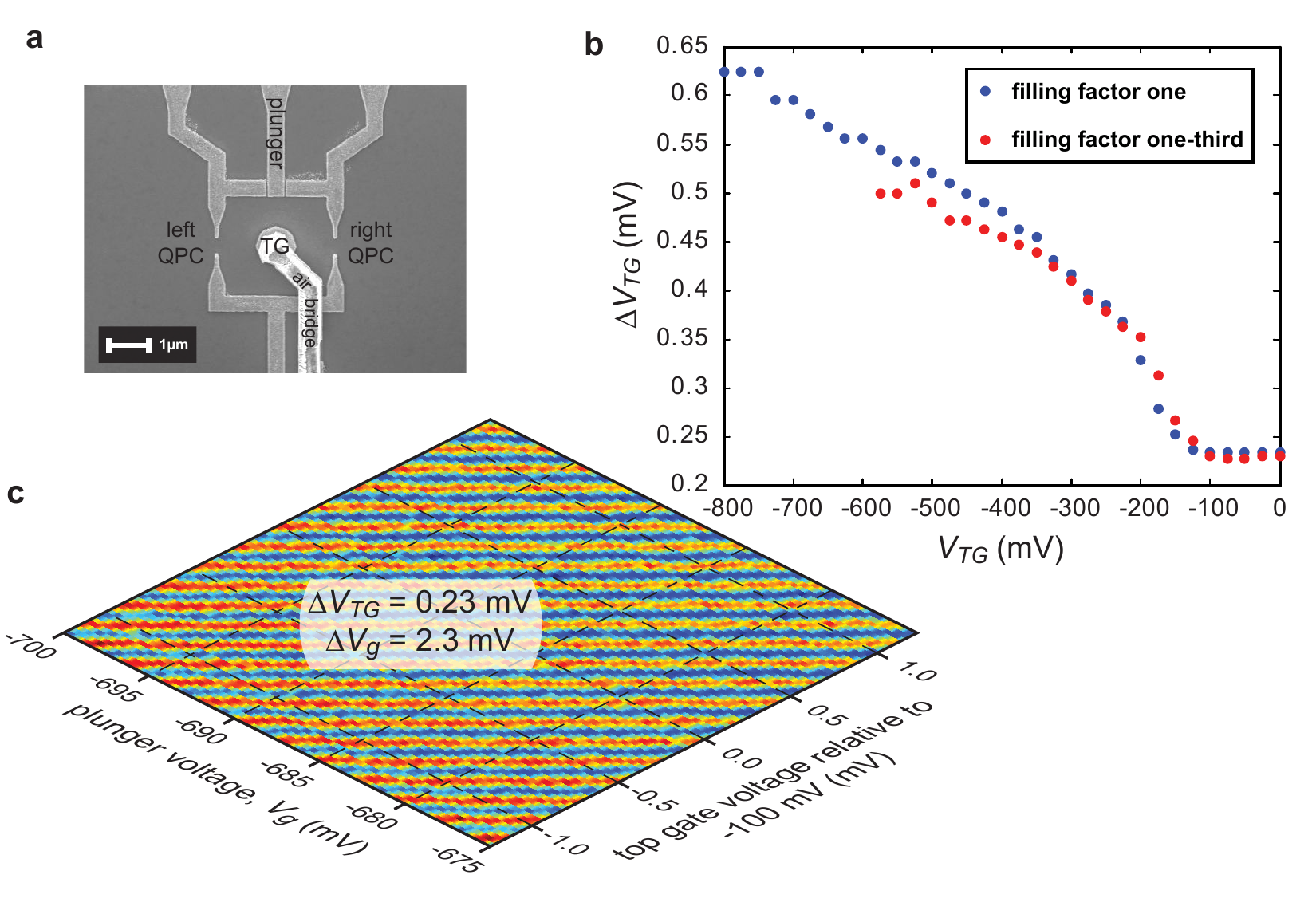}}
\caption{Interference controlled by a voltage applied to a `small top-gate' (deposited in the center of the FPI cell) and the plunger gate voltage.  (a) SEM picture of the FPI with a small top-gate.  The brighter areas are metallic gates.  (b) Periodicities of interference for $\nu$=1 and $\nu=1/3$ as function of gate voltage, before and after depletion under the gate.  (c)  Conductance in the plane of \VTG-\VG.  Note that the two gates function in a similar fashion, namely, depleting the area under (and around) them leads to a decrease in the area of the interferometer.}
\end{figure}

\begin{figure}[h!]
\label{fig:one_third}
\centerline{\includegraphics*[viewport=1.1in 1.8in 7.1in 9.55in,width=82mm]{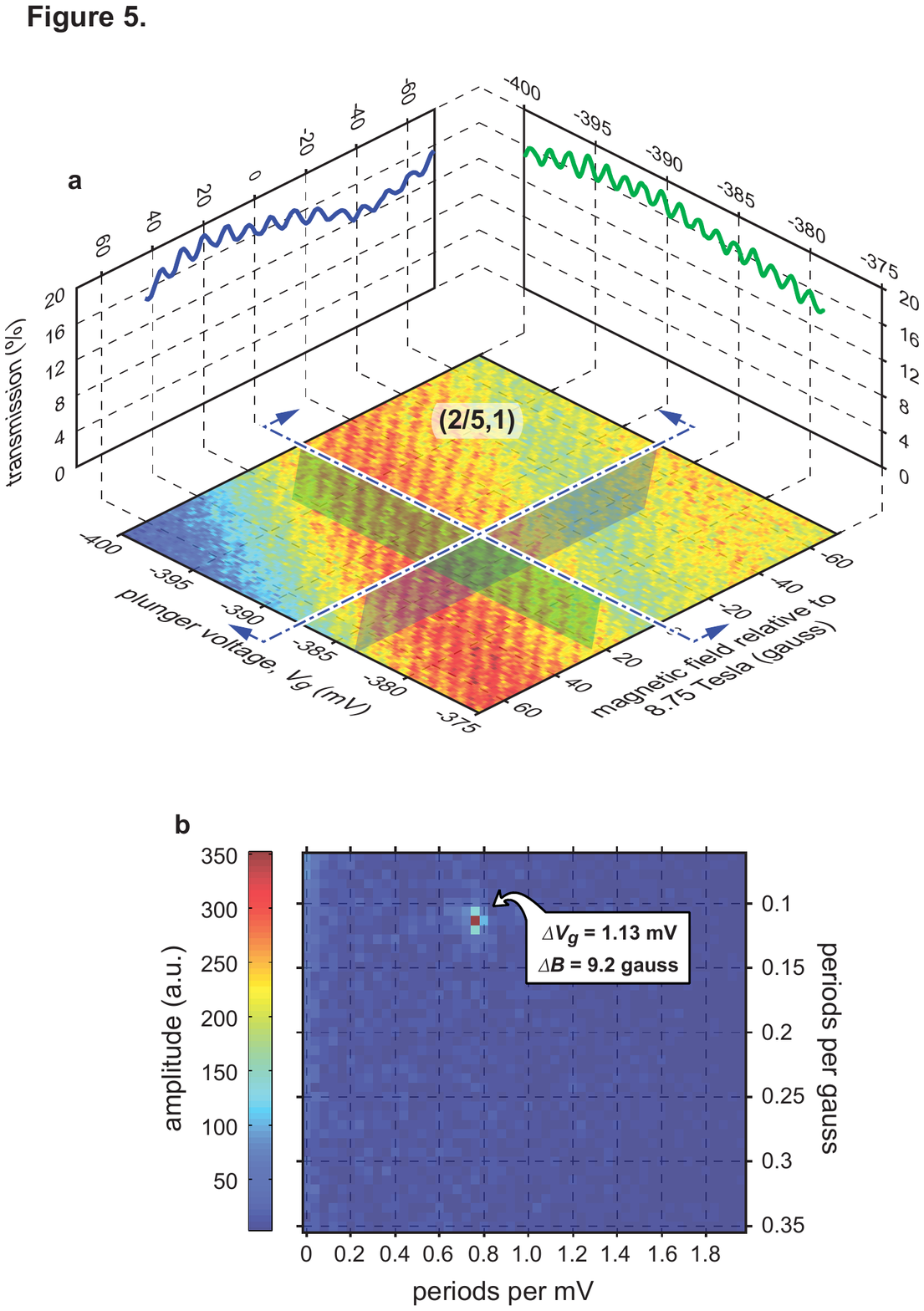}}
\caption{Interference of the inner channel for bulk filling factor $\nu=2/5$; case $(2/5, 1)$.  (a) Conductance plot in the $B$-\VG~plane for the inner channel.  Upper left and right panes shows traces with respect to $B$ and \VG~averaged over several traces.  (b) 2D FFT analysis of the interference oscillations.  The periodicity in \VG, being $1/3$ of that needed for expulsion of an electron, suggests the expulsion of $e/3$ charge per period, while the interfering quasiparticle is $e/5$ (see text).}
\end{figure}

We now discuss the FQHE regime, where interference was measured for $\nu$=4/3, $\nu$=2/3, $\nu$=2/5 and $\nu$=1/3.  Due to space limitation we discuss here only the latter two cases.  Like in the IQHE, no interference oscillations were observed as function of $B$ for cases with \fT=0, namely, at $(2/5, 0)$ and $(1/3, 0)$.  However, periodic oscillations of the conductance, as function of both $B$ and \VG, were observed at $(2/5, 1)$ as shown in Fig. \ref{fig:one_third}a.  While the quality of the data deteriorated somewhat, a fast Fourier transform of the bare data shows a single, reproducible, distinct peak, corresponding to $\Delta B$=9.2gauss and $\Delta$\VG=1.13mV (see Fig. \ref{fig:one_third}b).  We return to the analysis of this periodicity in the discussion section.

The results discussed so far can be encapsulated as follows:
\begin{enumerate}
  \item In each filling factor, any of the edge channels can be made to interfere independently of the others.
  \item The periodicities $\Delta B$ and $\Delta$\VG~do not directly depend on the magnetic field (or the filling factor) nor the transmission coefficients $t_1$ and $t_2$, but only on \fT.  The periodicity in magnetic field is solely determined by A and \fT:
\begin{equation}
\label{eq:b_periodicity}
\Delta B\left(B, V_g, f_T, t_1, t_2\right)=\Delta B\left(\mbox{\fT}\right)=-\frac{\phi_0}{A\cdot\mbox{\fT}},
\end{equation}
as predicted in Ref. \cite{ROS2007} for the case $t_1$, $t_2\ll1$; here, however, we found it to hold for all values of the transmissions.  Similarly, the periodicity in plunger gate voltage depends only on \fT:
\begin{equation}
\label{eq:vg_periodicity}
\Delta \mbox{\VG}\left(B, V_g, f_T, t_1, t_2\right)=\Delta \mbox{\VG}\left(\mbox{\fT}\right),
\end{equation}
with $\Delta$\VG~monotonically increasing with increasing \fT~(aside from two exceptions to be discussed below).  Fig. 6 summarizes the measured periodicities $\Delta B$ and $\Delta$\VG~in cases of IQHE and a single case of FQHE;
\item For \fT=0, the enclosed flux is independent of the magnetic field, indicating that the relevant area of the interferometer is proportional to $1/B$;
\item For \fT$>$0, the enclosed flux decreases with increased magnetic field, indicating that the interferometer area shrinks faster than $1/B$ (hence, the minus sign in Eq. \ref{eq:b_periodicity}).
\end{enumerate}

\begin{figure}[h!]
\label{fig:summary}
\centerline{\includegraphics*[viewport=1.1in 3.4in 7.1in 8.1in,width=82mm]{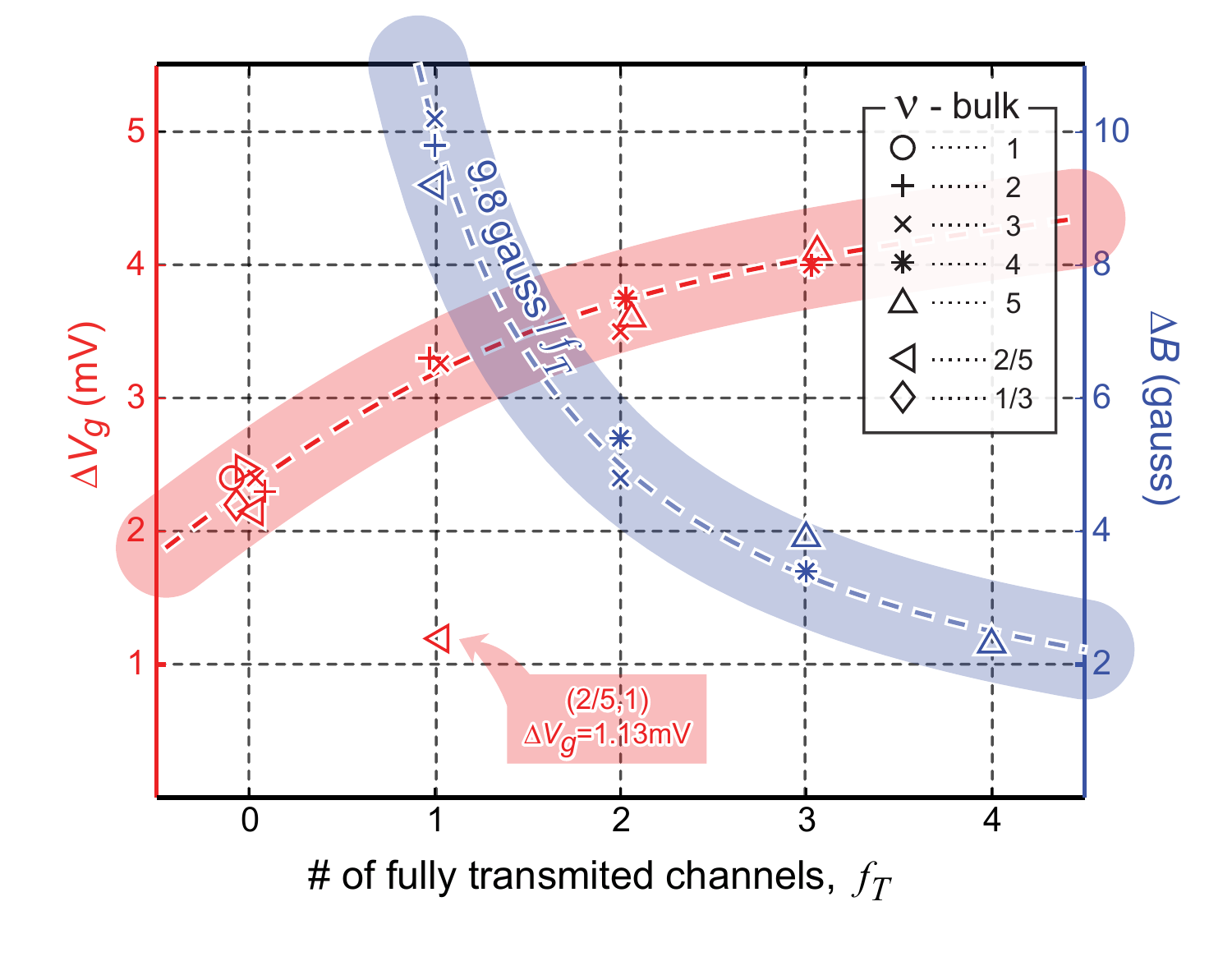}}
\caption{Summary of the periodicities.  In red (left axis) - periodicities in \VG~as a function \fT, for different filling factors.  A larger number of freely propagating outer channels screens the gate more efficiently, thus, increasing the periodicities.  In gray (right axis) - periodicities in B as a function of \fT.  The periodicities are inversely proportional to \fT.}
\end{figure}

The first, and most important, exception to these general statements is the periodicity $\Delta$\VG~in the case $(2/5, 1)$: it is $1/3$ of the measured value for all other configurations with \fT=1.  The second exception takes place for high filling factors, when a large number of channels are being trapped; such as $(4, 0), (4, 1), (5, 0)$.  As seen in the example $(4, 1)$ in Fig. 7, a `lattice-like' structure in the $B$-\VG~plane appears instead of the more ubiquitous `pyjama-like' stripes.  The dependence on $B$ and \VG~is still periodic, but it cannot be described by a periodic function of a linear combination of $B$ and \VG.  This behavior is not yet understood.

\begin{figure}[h!]
\label{fig:exotic}
\centerline{\includegraphics*[viewport=1.1in 3.5in 7.1in 8.0in,width=82mm]{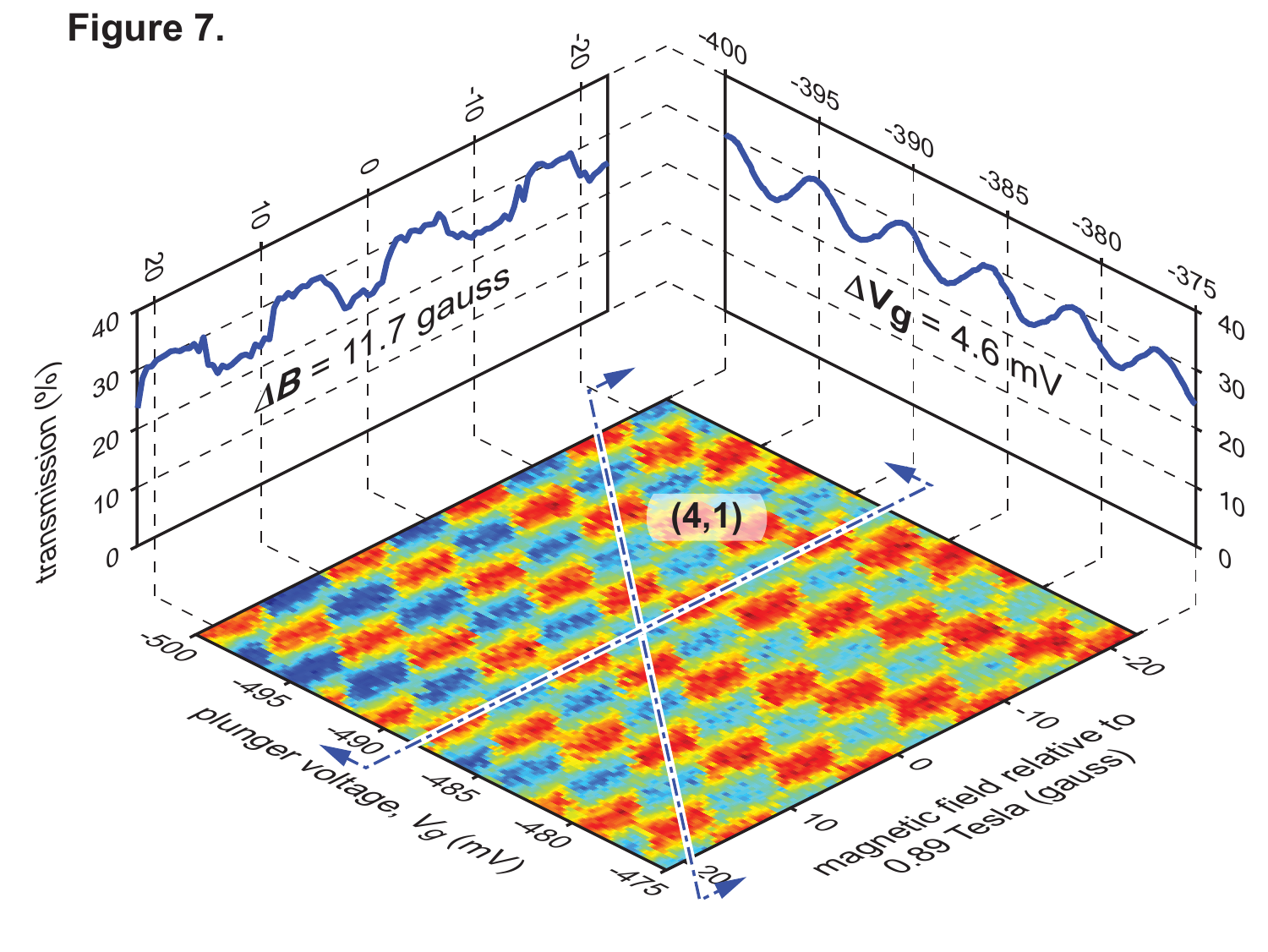}}
\caption{An exotic behavior of the oscillations for case $(4, 1)$, as an example of a typical behavior of many trapped channels in the FPI cell area.  Here, there is a sinusoidal dependence on \VG~(upper right pane), however, modulated by a square-type pulse train.  Scanning $B$ while keeping \VG~fixed produces a sharp pulse train (upper left pane).  Other cases, like $(5, 0)$ and $(6, 0)$, produce more complicated patterns.}
\end{figure}

\section{Discussions}

\subsection{Magnetic field periodicity}
The empirical Eqs. \ref{eq:b_periodicity} and \ref{eq:vg_periodicity} may be understood by means of a single physical assumption: the relevant area of the interferometer varies as the magnetic field is varied, namely, the magnetic flux through FPI area is $\Phi=B\cdot A(B, V_g)$, with A depending on \VG~and $B$.  Then, the phase evolves as function of $B$ and \VG:
\begin{eqnarray}
\nonumber d\Phi&=&\frac{\partial(A\cdot B)}{\partial B}dB+\frac{\partial(A\cdot B)}{\partial V_g}dV_g\\
&=&\left(A+B\frac{\partial A}{\partial B}\right)dB+B\frac{\partial A}{\partial V_g}dV_g,
\end{eqnarray}
with the lines of constant phase (constant flux) in the $B$-\VG~plane given by:
\begin{equation}
\frac{dB}{dV_g}=-{\frac{B}{A}\frac{\partial A}{\partial V_g}}\left/{\left(1+\frac{B}{A}\frac{\partial A}{\partial B}\right)}\right..
\end{equation}

In earlier experiments with an electronic two-path MZI, we found that ${\partial A}\left/{\partial B}\right.\approx0$, with the lines of constant phase having an opposite slope from the one measured with the FPI \cite{NED2007,JI2003}.  Assume that all \fT~fully transmitted channels enclose a similar area $A_T$.  This area is expected to be field independent since the channels are in equilibrium with the leads.  The inner channels are trapped and thus maintain a discrete and constant number of electrons $N_R$ for small changes in $B$, while a smooth electron exchange can take place between the interfering channel and the fully transmitted ones.  The total amount of charge in the interferometer cell is
 $Q=e\left[N_R+\frac{B}{\phi_0}\left(f_TA_T+A\right)\right]$.  If charge neutrality is a major consideration then $Q$ should be independent of $B$, namely, $\partial Q\left/\partial B\right.=0$ .  Assuming $A\sim A_T$,
\begin{equation}
\frac{\partial (AB)}{\partial B}=-f_T\cdot A,
\end{equation}
implying that the flux, and hence the AB phase, indeed \emph{decreased} with increasing magnetic field at a rate proportional to \fT, and were independent of $B$ for \fT=0; as observed in both the integer and fractional cases.

For \fT=0, increasing $B$ is followed by a decrease in the area $A$ in such a way that the flux, and hence the number of occupied states, are kept constant, but the charge density at the center of the FPI cell grows.  A spatial imbalance between electrons and ionized donors takes place.  Increasing the field further must lead eventually to relaxation of that imbalance by creation of a hole within the interfering Landau level if this is also the only edge channel, or electron tunneling from the trapped channels otherwise. In any case, this leads to abrupt increase of one quantum flux to the interfering loop (hence, invisible in the interference pattern). The area `breathes' with magnetic field. It decreases monotonically and increases abruptly while keeping the average constant. The same holds for \fT$>$0. The only difference is that the area decreases with increasing $B$ even faster, since the interfering channel loses charge monotonically to the \fT~lower filled Landau levels (whose density increases with field) leading to the positive slope of the constant phase lines in the $B$-\VG~plane.

\subsection{Gate voltage periodicity}
The periodicity in gate voltage $\Delta$\VG, and in particular its relation to the fractional filling factor $\nu$, provide an insight to the charge of the interfering quasiparticle.  The dependence on \VG~can be understood if we assume that the capacitance $C$ between the plunger gate and the interfering channel depends only on the number of fully transmitted channels (that screen the plunger gate voltage), namely, $C=C(f_T)$.  The interfering channel flows at the interface between two areas with different filling factors, $\nu_{out}$ and $\nu_{in}$, each with charge $q_i$ per flux quantum $q_i=e\nu_i$.  As long as the sole function of biasing the plunger gate by $\delta V_g$ is to move the interface between the two filling factors by area $\delta A$, and thus expel charge $\delta q$ from within the interferometer, we may write:
\begin{equation} \label{eq:capacitance}
  \delta q=C\delta V_g=\frac{B\cdot\delta A}{\phi_0}\cdot e\left(\nu_{in}-\nu_{out}\right).
\end{equation}
Assuming that the change in area does not change the number of quasiparticles enclosed by the interfering loop (as their number is very small near the center of the conductance plateau), the change of area leads to a change in the AB phase,
\begin{equation}
  \delta\phi=2\pi\frac{e^*}{e}\frac{B\cdot\delta A}{\phi_0}.
\end{equation}
Using Eq. \ref{eq:capacitance} we get a relation between the interfering quasiparticle charge $e^*$ and the periodicity $\Delta$\VG:
\begin{equation}\label{eq:charge}
  \frac{e^*}{e}=\frac{\nu_{in}-\nu_{out}}{\Delta V_g/V_e},
\end{equation}
with $V_e\equiv e/C$ the gate voltage needed to expel one electron.

Look first at cases $(1/3, 0)$ and $(2/5, 0)$; both with a gate voltage periodicity $\Delta$\VG$\sim$2.2mV (which is the similar value as for \fT=0 in the integer cases).  Evidently, with the assumption $C=C(f_T)$, with $C$ being independent of the filling factor, the measured expelled charge per period must be in both cases $e$.  Since the interfering edge channel in both cases belongs to the $1/3$ fractional state, $\nu_{in}-\nu_{out}=1/3$, and the interfering quasiparticle charge, deduced directly from Eq. \ref{eq:charge}, must be $e^*=e/3$; agreeing with our expectations.

What happens in the case $(2/5, 1)$, where the interfering channel belongs to the $2/5$ fractional state?  The observed periodicity is $\Delta$\VG$\sim$1.13mV (see Fig. 5) - nearly $1/3$ of the period observed in the integer cases $(\nu , 1)$.  Since the capacitance $C(f_T=1)$ was found to be independent of filling factor, and the measured periodicity is a third of that observed in the integer regime, the expelled charge per period of the oscillation must be $e/3$ (directly proportional to $\Delta$\VG).  Again, for $\nu_{in}-\nu_{out}=1/15$ in this case, and the above periodicity, the interfering charge is $e^*=e/5$.  This is a striking example of an expelled charge $e/3$ per period of gate voltage, while the interfering quasiparticles carried charge $e/5$.

Since the periodicity of the oscillations is independent of $t_1$ and $t_2$, one can invoke a simpler explanation \cite{ROS2007} for the expelled charge in each period, which is valid only in the Coulomb blockaded limit, namely, when $t_1$ \& $t_2\ll1$.  In that limit, the charge enclosed by the constrictions must be quantized in units of the elementary charge in the incompressible regions within the two constrictions.  For cases $(1/3, 0)$ and $(2/5, 0)$, the filling factor within the constrictions is zero and the expelled charge is that of the electron; as indeed was observed.  Alternatively, for $(2/5, 1)$, the filling factor in the constrictions is $1/3$ and the expelled charge must be $e/3$; again, agreeing with the observation.

\begin{figure}[h!]
\label{fig:top_gate}
\centerline{\includegraphics*[viewport=1.2in 1.5in 7.1in 9.5in,width=82mm]{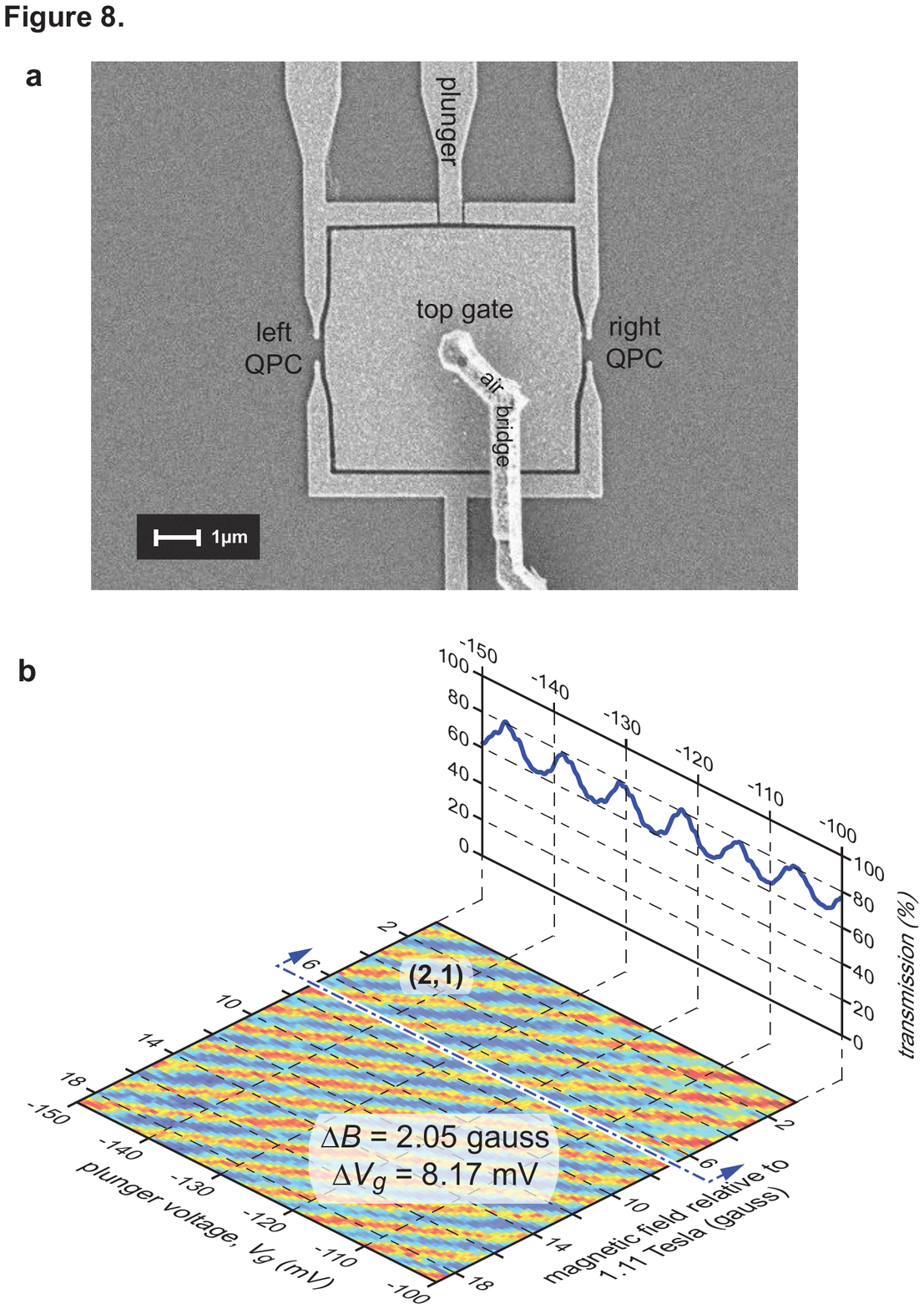}}
\caption{The behavior of a FPI covered by a grounded top-gate covering the full area of the interferometer.  (a) SEM image of the larger FPI (active area $\sim$17$\mu$m$^2$) with the top-gate.  (b) Conductance in the $B$-\VG~plane for case $(2, 1)$ exhibiting an opposite slope of the constant phase lines in the uncovered interferometer.}
\end{figure}

\subsection{Top gated device}

 The significance of Coulomb interactions in the FPI was tested by adding a grounded metal gate that fully covered the FPI's area (`large top-gate', or LTG); as was also done in Ref. \cite{ZHA2009}.  This gate effectively increased the screening of the uncompensated charges.  While the addition of the LTG to the smaller FPI (area 4.4$\mu$m$^2$) did not affect the orientation of the constant phase lines (it did increase the periodicity in plunger gate voltage, which is expected as the plunger gate is less effective due to screening provided by the LTG biasing), adding a LTG to the larger FPI (area 17.6$\mu$m$^2$, as seen in Fig. 8a) led to markedly different results.  First, unlike ALL our previous results, AB oscillations were measured as function of magnetic field also for cases with \fT=0, such as $(2, 0)$ \& $(1, 0)$.  Second, as seen in Fig. 8b for the case $(2, 1)$, the orientation of the constant phase lines in the $B$-\VG~plane was reversed, indicating an increased flux with increasing magnetic field (see also \cite{ZHA2009}).  However, the area determined from the AB periodicities in $B$ still did not match the expected area.

\section{Conclusion}
Our experiments provide a detailed first step study towards the realization of interferometers that might give direct evidence of the statistics of fractionally charged quasiparticles.  Our studies covered interference of the different edge channels of complex edge structures in integer and a few fractional states in the QHE regime.  We found that the interferometers are very strongly affected by Coulomb interactions, which radically modify the interferometer area with magnetic field - independent of the coupling strength to the leads (in or away from the Coulomb blockade regime).  Interfering charges of $1e$, $e/3$ and $e/5$ where deduced from the interference patterns.

\section{acknowledgments}
We are grateful to B. I. Halperin, C. Marcus, I. Neder, B. Rosenow, A. Yacoby and Y. M. Zhang for helpful discussions.  We acknowledge the partial support of the Israeli Science Foundation (ISF), the Minerva foundation, the German Israeli Foundation (GIF), the German Israeli Project Cooperation (DIP), the European Research Council under the European Community's Seventh Framework Program (FP7/2007-2013)/ERC Grant agreement \# 227716, the US-Israel Bi-National Science Foundation, and Microsoft's Station Q. N. Ofek acknowledges support from the Israeli Ministry of Science and Technology.

\end{document}